# Electron-Backscattering-Assisted High Harmonic Generation from Bilayer Nanostructures

Chao Yu,[1] Shicheng Jiang,[1] Tong Wu,[1] Guanglu Yuan,[1] Yigeng Peng,[1] Cheng Jin,[1] and Ruifeng Lu[1,2,*]

[1]*Institute of Ultrafast Optical Physics, Department of Applied Physics, Nanjing University of Science and Technology, Nanjing 210094, P R China*

[2] *State Key Laboratory of Molecular Reaction Dynamics, Dalian Institute of Chemical Physics, Chinese Academy of Sciences, Dalian 116023, P R China*

In the framework of time-dependent density functional theory, we obtain high-order harmonics of photon energies up to 10 $U_\mathrm{p}$ from bilayer crystals with an interlayer spacing $d = 70$ Å. At grazing incidence, a clear double-plateau structure is observed in the harmonic spectrum. The photon energy of the second plateau far beyond atomic-like harmonics can be well explained by the inclusion of backscattering of ionized electrons. Ab initio simulations reveal that the cutoff of the second plateau is continuously extended with an increasing $d$. Our classical calculations predict that the maximum electronic kinetic energy is linearly dependent on $d$ over a wide range. Moreover, the harmonic yield in the second plateau is significantly enhanced by increases in the wavelength of the driving laser. Owing to the confined spreading of the electronic wave packet, a beneficial wavelength scaling of $\lambda^{2.85}$ is obtained. This study therefore establishes a novel and efficient way of producing high-energy light source based on layered nanostructures.

---

[*] rflu@njust.edu.cn

High-order harmonic generation (HHG) is one of the fundamental processes in intense laser–matter interaction and is regarded as one of the most prominent methods for producing attosecond light pulses. In the past decades, HHG from gaseous media has been extensively investigated [1–5], however two key ,challenges for future HHG experiments remain to be resolved, which are as follows: (i) continuously increasing the harmonic conversion efficiency and (ii) extending the cutoff energy to achieve practical attosecond photonics [6].

Recently, the possibility of realizing efficient high harmonics from solid materials has attracted a considerable amount of attention [7–17], as it offers opportunities for obtaining extreme ultraviolet sources [11, 18–20] and studying ultrafast electron dynamics in condensed matter [21, 22]. By applying a sub-cycle synthesized field, Luu et al. [11] experimentally determined that the total photon yield of fused silica is higher than those of noble gases. In the search for highly efficient solid media for HHG, two-dimensional (2D) materials, including graphene [13, 23–25], monolayer molybdenum disulfide ($MoS_2$) [13, 26], and hexagonal boron nitride (h-BN) [27–29], have become the focus of various studies regarding this topic. As observed in experiments for $MoS_2$ [26] and predicted in theory for h-BN [27], monolayer materials are promising for efficient HHG owing to their unique electronic properties, which are lacking in the corresponding bulks.

In solid HHG, the energy cutoff scales linearly to the peak field [7] and also depends strongly on the electronic band structure [11, 30]. However, the maximum intensity that could be applied to a bulk material is intrinsically limited by its damage threshold. Hence, the cutoff of solid HHG is, in principle, restricted by the peak intensity of the driving laser and the maximum energy gap at the Brillouin zone edge [31, 32]. Unfortunately, past experiments have also demonstrated that the cutoff in a solid harmonic spectrum is unfavorable compared to that for a gas, which can reach the soft X-ray range [16]. More recently, Tancogne-Dejean et al. [33] demonstrated atomic-like HHG in monolayer h-BN when laser polarization is perpendicular to the material surface. Indeed, this is an interesting novel approach for generating high-order harmonics based on a solid-state device, with an energy cutoff and a more favorable wavelength scaling of the harmonic yield similar to those in atomic and molecular gases.

Because atomic-like HHG in the monolayer is characterized by advantages both in terms of high cutoff energy, typically exhibited in gases, and high conversion efficiency, typically exhibited in solids [33], we extend the 2D target to bilayer geometry to explore the interlayer scattering in solid HHG. In this letter, we consider h-BN as the prototypical case. The bilayer h-BN is constructed with large interlayer spacing in the order of nanometers. In practical experiments, such h-BN bilayer nanostructures with large interlayer spacing can be fabricated via physical assembly using spacers [34, 35] or a chemical etching method [36]. We do not aim to discuss interlayer coupling inside bilayer with equilibrium distances less than 4 Å, which has already been addressed to a certain extent in previous research [27]. In fact, the stacking pattern and ingredients of a distant bilayer will not affect the main conclusions. Furthermore, highly complicated infrastructure analogous to that of a metasurface [37] is not within the scope of this study.

To drive the ionization of electrons populating the valence band of h-BN without causing damage, we use strong laser grazing incident to the material surface. In the framework of the time-dependent density functional theory (TDDFT), the HHG spectra of monolayer h-BN and distant bilayer h-BN are simulated using the Octopus code [38]. In the previous works, TDDFT has been used to study the ellipticity dependence of HHG in bulk silicon and MgO [39], the polarization-state-resolved high-harmonic spectroscopy from silicon [40], and HHG in some complex systems, such as strongly correlated material [41] and spin-polarized h-BN with defects [42]. We include a full description of electron–electron and electron–ion interactions in the TDDFT calculations, and additional numerical details are provided in the supplemental materials.

In Fig. 1, the simulated high harmonic spectra from a monolayer and h-BN bilayers with interlayer spacings of $d = 40$ Å (4 nm) and $d = 70$ Å (7 nm), are displayed. We select almost the same laser parameters as in Ref. 33. The calculated work function $E_w$ of monolayer h-BN is approximately 6.0 eV based on ground state calculation within the local density approximation, thereby suggesting that layered h-BN can be exposed to such an intense and short laser for an out-of-plane electric field without undergoing damage [33]. In addition, under such laser parameters, Zener tunneling dominates the transition according to Keldysh parameter $\gamma = \sqrt{\dfrac{E_w}{2U_p}}$, where $U_p = I/4\omega^2$ is the ponderomotive energy of the electron in the applied laser field. For the monolayer h-BN, the emitted harmonics exhibit atomic-like HHG features [33], and the cutoff energy, as presented in Fig. 1 (green line), is in good agreement with the value ($E_{cutoff} = E_w + 3.17\ U_p$) predicted by the three-step model [43, 44].

In Fig. 2(a), the physical processes in the traditional three-step model are schematically plotted for HHG from the monolayer h-BN. First, electrons are released via tunneling near each peak of the electric field. In a periodically oscillating laser field, the electrons are driven to the farthest point where their velocity becomes zero at roughly half an optical cycle. Electrons can then be accelerated by the laser field in the reverse direction and returned to the monolayer h-BN. Finally, they recombine with the parental material, thereby releasing the energy acquired from acceleration in the form of photon emission.

Significantly different from that of the monolayer case, the harmonic spectra of the bilayer h-BN with a large interlayer spacing present a distinct double-plateau profile, as depicted in Fig. 1, with a blue line for $d = 40$ Å and red line for $d = 70$ Å. Apparently, the cutoff energy values corresponding to the first plateaus for both $d = 40$ Å and $d = 70$ Å are almost the same as that of the monolayer, which indicates that the atomic-like three-step model depicted in Fig. 2(a) also dominates the harmonic emission in the first plateau. However, the second plateau, with photon energy far beyond that of the atomic-like HHG, could not be interpreted by the three-step model. Ab initio simulation results show that the cutoff values for the second plateaus are significantly extended to $E_w + 4.89\ U_p$ and $E_w + 9.53\ U_p$ for $d = 40$ Å and $d = 70$ Å, respectively.

To understand these extraordinary extensions related to the second plateau, a four-step model, wherein backscattering of the electrons from the neighbor layer leads to recombination at the parent layer, is illustrated in Fig. 2(b): (i) electrons are released from each h-BN monolayer; (ii) as the electronic movement is confined by the interlayer space, the electrons will be backscattered when driven from one layer to the other; (iii) electrons are accelerated by the laser field in reverse direction and return with a nonzero initial velocity after backscattering; and (iv) electrons recombine with the parental layer, releasing photons of higher energy. Notably, the first and final steps (i.e., electrons releasing and recombining) in the four-step model are the same as those in the traditional three-step model. In fact, the four-step model is analogous to the pioneering model developed for ultrahigh-order harmonic generation in ion–atom collisions [45] and two-center molecules at large internuclear distances [46]. However, this model is difficult to realize in experiments involving gaseous diatomic systems, because atoms or ions placed at random positions around the laser-driven atom are not sufficient for extending the harmonic plateau, and random variations in the harmonic phase destroy the coherence of emission [46]. For bilayer crystals, atoms are regularly arrayed with a high density in the crystal lattice; this is beneficial to the coherence of ultrahigh-order harmonic emission. Normally, a maximum energy of 10 $U_p$ for electrons can be achieved when electrons are scattered by 180 °[47]. Therefore, the additional backscattering step that endows electrons with large initial velocities for the following acceleration is key to forming a second, high-energy plateau.

For a more intuitive description of the backscattering process, the time evolution of the induced electronic density for distant bilayer h-BN is depicted in Fig. 2(c). The interlayer spacing here is $d = 40$ Å, and the laser parameters are the same as those in Fig. 1. Clearly, the time-dependent electronic wave packets evolve in bundles with two types of visible trajectories that return to the parental layer and result in the HHG. Specifically speaking, the trajectories labeled with purple hollow arrows outside the bilayer represent the processes in the traditional three-step model, whereas the brand-new trajectories marked with white hollow arrows between the two layers are part of the four-step process. As predicted, both types of trajectories emerge periodically every half optical cycle, thereby indicating that trajectories either in the three-step model or four-step model lead to harmonic emission at every half optical cycle.

To study the harmonic spectra in the time domain, time-frequency analyses of HHG from bilayer h-BN with $d = 40$ Å and $d = 70$ Å are presented in Fig. 3(a) and 3(b), respectively. As can be seen from Fig. 3(a), the harmonic spectra in the time domain comprise two plateaus: one that originated from the three-step model, with a maximum energy of approximately $E_w+3.17\ U_p$, and the other due to the four-step process, including backscattering, with a cutoff reaching $E_w+4.89\ U_p$. For multiple optical cycles, several central peaks in the time-frequency maps corresponding to the second plateau are labeled as $P_1$, $P_2$, $P_3$, and $P_4$ for convenience. Similar to the harmonic emission features in the first plateau, harmonics also burst at every half optical cycle in the second plateau; however, the emission time of each peak in the second plateau is shifted earlier than in the first plateau, which can be explained by the motion constraint of ionized electron inside the interlayer region.

To better elucidate the striking phenomena derived from the TDDFT quantum calculations, we develop a simplified model based on classical Newtonian equations. In this model, the electronic motion in the aforementioned laser field $E_L(t)$ can be described by $\ddot{x} = -E_L(t)$. Here, the Coulomb attraction to the parent layer is neglected, and we assume that the electron emerges with a zero velocity, $\upsilon(t_0) = 0$, on the outside of the tunneling barrier. Thus, the velocity of the electron following the laser field can be described by $\upsilon(t) = A_L(t) - A_L(t_0)$, where the vector potential $A_L(t)$ is defined by $E_L(t) = -A_L(t)/dt$. Backscattering occurs once the electrons arrive at the other layer. We note that the backscattering calculated here is regarded as an elastic scattering, which implies that the momentum of the electron is reversed into $\upsilon_b(t) = -\upsilon(t)$ during the backscattering. We focus only on the electrons that indicate ionization and recollision in one optical cycle, because seriously spreading the electron wave packet over a long time could heavily reduce the signals of HHG. The classical ionization and recollision energy diagrams for bilayer h-BN with $d = 40$ Å and $d = 70$ Å are shown in Fig. 3(c) and 3(d), respectively. In Fig. 3(c), we clearly see that both long and short trajectories contribute to HHG in the first plateau, which is consistent with the time-frequency analyses in Fig. 3(a). It is well established that a free electron can gain a maximum energy of 3.17 $U_p$, a limit that obeys the cutoff position determined in the first plateau. On the other hand, the origin of the second plateau in HHG from bilayer h-BN with large interlayer spacing can be analyzed perfectly using our classical model, which accurately computes an energy value of 4.89 $U_p$ based on a momentum reversion after backscattering. Moreover, three trajectories in Fig. 3(c) contribute to HHG in the second plateau, agreeing well with the time–frequency analysis in Fig. 3(a). Because the trajectories ($T_r$ **1** and $T_r$ **2**) will degenerate when the interlayer spacing is increased (see Fig. s1), we define the maximum energy of trajectory **3** ($T_r$ **3**) as the cutoff energy. Compared to the case for $d = 40$ Å, the electron for $d = 70$ Å can achieve an energy as high as 9.53 $U_p$ in the second plateau (Fig. 3(d)).

Furthermore, the dependence of HHG on the interlayer distance is illustrated in Fig. 4(a). Notably, the laser parameters are identical to those presented in Fig. 1. It is natural that the cutoff energy ($E_{cutoff} = E_w + 3.17\ U_p$) of the first plateau remains unchanged. For the second plateau, the purple line in Fig. 4(a) represents the cutoff values estimated using the classical model. The cutoff of the second plateau is extended nearly linearly with the increase in the interlayer spacing of the bilayer h-BN. However, the conversion efficiency of the second plateau is significantly weakened for larger interlayer spacings, which is understandable considering that the chosen laser condition allows less electrons to reach the other layer for $d > 70$ Å and that the spreading of the electron wave packet is affected by interlayer spacing. Furthermore, we determine the correlation of interlayer spacing $d$ with maximum kinetic energy of electron to be $E_k = 5.185\ U_p d/\alpha$ by fitting the results from classical calculation in the range $1.0 < d/\alpha < 4.0$, where $\alpha = E_0/\omega^2$ is the quiver amplitude of an electron driven by a laser field. As shown in Fig. 4(b), the maximum classical kinetic energy approaches the atomic limit (3.17 $U_p$) for $d \rightarrow$ equilibrium distance (~3.4 Å). If $d \rightarrow$ infinity (see Fig. s2), the bilayer system can be regarded as two non-connected single layers, such that electrons ionized from one layer can hardly reach the other layer, and thus the maximum classical kinetic energy approaches the atomic limit (3.17$U_p$).

In terms of wavelength dependence, we simulate the HHG from a distant bilayer with interlayer spacing fixed at $d = 70$ Å, as shown in Fig. 4(c). For the HHG from noble gas atoms, theoretical calculations predict that harmonic yield falls dramatically with a very unfavorable $\lambda^{-(5-6)}$ wavelength scaling [48, 49] and even worse in experiments [50]. In Fig. 4(c), the atomic-like harmonic yield in the first plateau decreases drastically as the laser wavelength is increased, because of a time–space spreading of the electron wave packet during its free evolution in vacuum [51, 52]. Nevertheless, in Fig. 4(d), the integrated harmonic yield from 150 to 200 eV in Fig. 4(c) exhibits a benign wavelength scaling law, i.e., $\lambda^{2.85}$, in the second plateau. Because a longer wavelength will permit slow electrons to move further in space, accordingly more electrons can travel from one layer to the other and are backscattered. Moreover, as shown in the classical ionization and recollision plots in Fig. s3, trajectories **1** and **2** gradually survive. Therefore, the harmonic yield of the second plateau is significantly enhanced with increases in the wavelength.

In conclusion, from ab initio TDDFT calculations together with quantum time–frequency and classical model analyses, we suggest a not-yet-discovered strategy for achieving high-intensity and high-energy harmonics from solids, i.e., with the advantages of electronic releasing and backscattering inside bilayer crystals characterized by interlayer spacing at the nanometer scale. At grazing incidence, laser–material interaction is anticipated to produce a second plateau with photon energies far beyond $E_w+3.17\ U_p$, and the harmonic yield is hopefully enhanced via adjustment of the interlayer spacing and wavelength. Importantly, the wavelength scaling of harmonics yield assisted by electron backscattering from distant bilayer nanostructures is superior to that from gaseous atoms. Last but not the least, essential features of HHG in the second plateau are noted to be not sensitive to the stacking pattern (see Fig. s4) and types (see Fig. s5) of bilayer materials. We expect that electron-backscattering-assisted HHG from a distant bilayer can be optimized in a multilayer array (see Fig. s6) and highly complicated nanostructures, which can be fabricated with the use of diverse nanoengineering technologies for practical application.

## ACKNOWLEDGEMENTS
This work was supported by NSF of China Grant (No. 11704187, 11974185, 11774175 and 11834004), the Natural Science Foundation of Jiangsu Province (Grant No. BK20170032), Fundamental Research Funds for the Central Universities (No.

30920021153) and the Project funded by China Postdoctoral Science Foundation (Grant No. 2019M661841). The Octopus code is available from www.octopus-code.org.

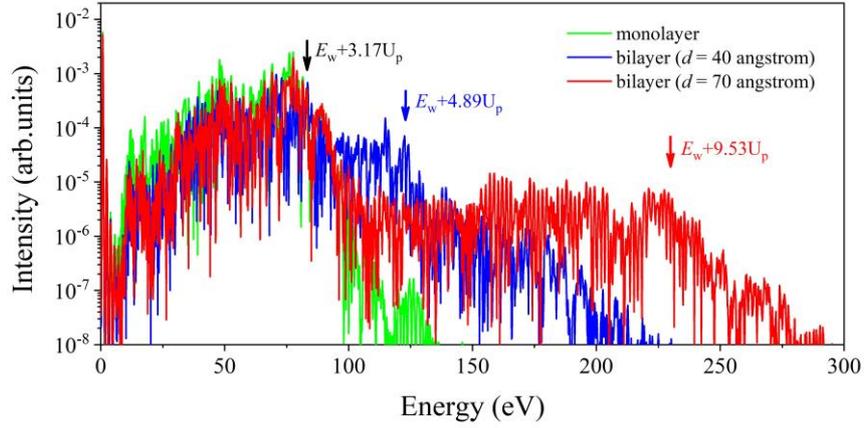

Fig. 1. High harmonic spectra from monolayer (green line), bilayer h-BN with interlayer spacing $d = 40$ Å (blue line) and $d = 70$ Å (red line). Harmonic yields for bilayer nanostructures are values per layer. Laser polarization is perpendicular to h-BN surface, and peak intensity is $I = 1.0 \times 10^{14}$ W/cm$^2$. Laser wavelength is 1600 nm, and full width at half maximum of Gaussian-envelop laser is 15 fs.

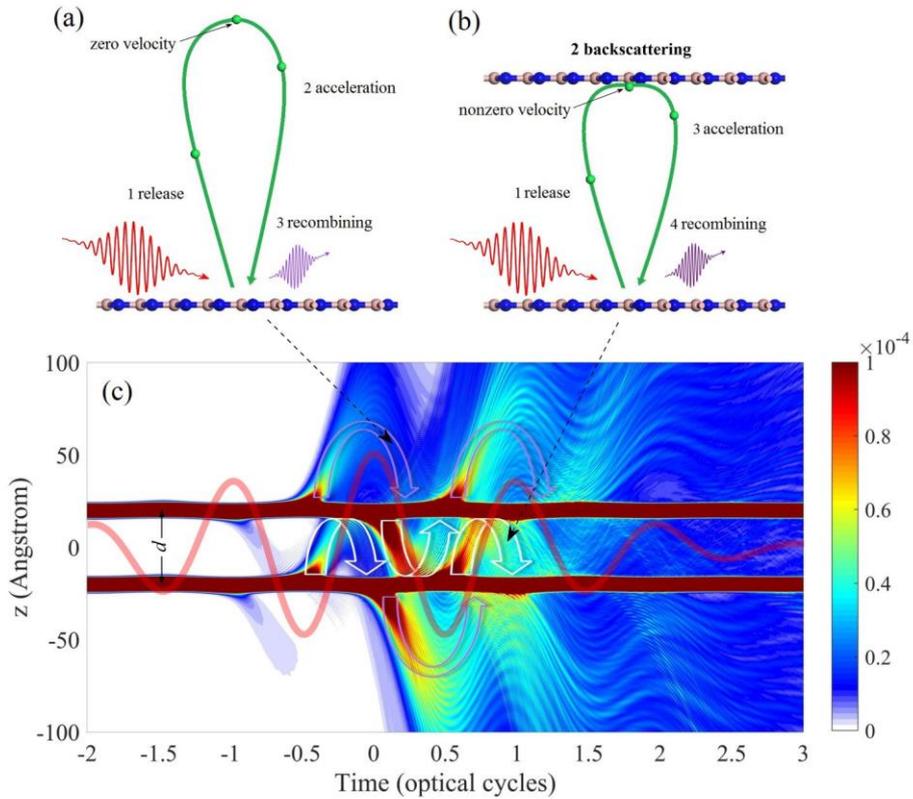

Fig. 2. Schematic HHG processes from (a) traditional three-step model in monolayer h-BN and (b) four-step model in distant bilayer h-BN. (c) Time evolution of induced electronic density for distant bilayer h-BN. Interlayer spacing is $d = 40$ Å, and laser parameters are same as those in Fig. 1.

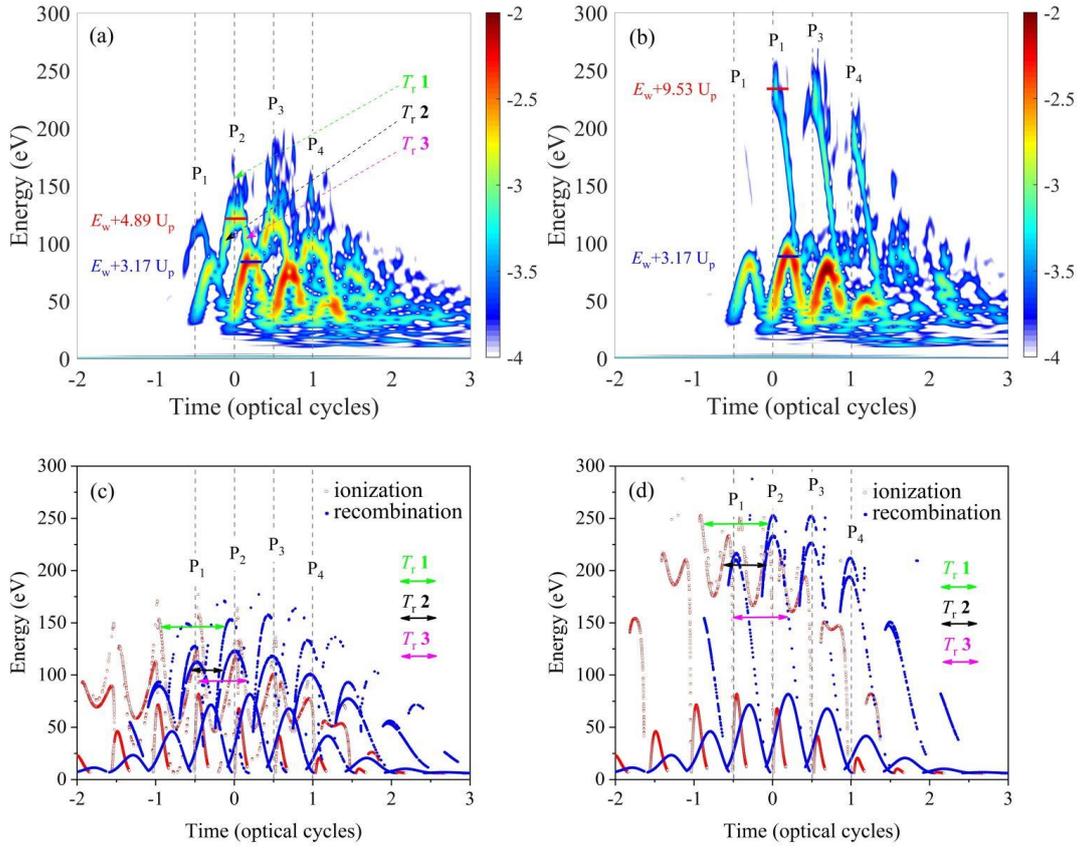

Fig. 3. Time-frequency analyses of HHG from distant bilayer h-BN for (a) $d = 40$ Å and (b) $d = 70$ Å. Classical ionization and recollision energy diagrams for distant bilayer h-BN with interlayer spacing of (c) $d = 40$ Å and (d) $d = 70$ Å. Laser parameters are same as those in Fig. 1.

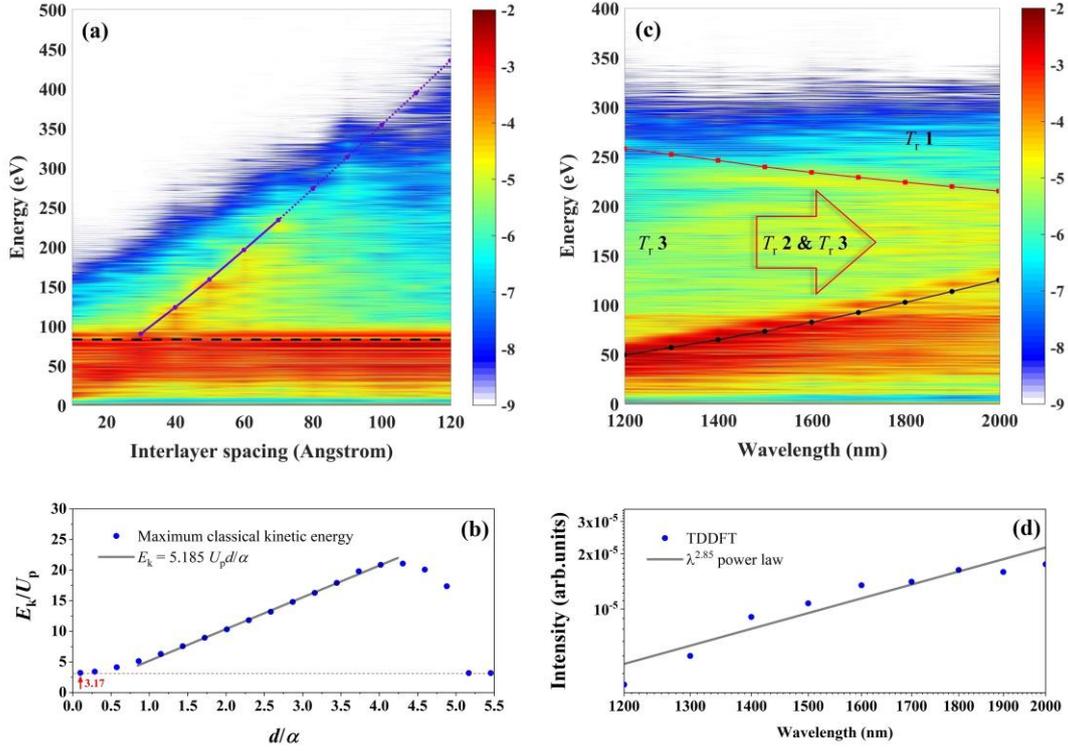

Fig. 4. (a) Interlayer-spacing-dependent HHG. Purple line represents harmonic cutoff calculated using classical model. Black dashed line represents cutoff in first plateau. Laser parameters are same as those in Fig. 1. (b) $d$-dependent maximum classical kinetic energy. (c) Wavelength-dependent HHG with fixed interlayer spacing $d = 70$ Å. Black line represents cutoff in first plateau, and red line represents cutoff of trajectory **3** ($T_r$ **3**) in second plateau. Except for wavelength, other laser parameters are identical to those in Fig. 1. (d) Harmonic yields integrated from 150 to 200 eV in (c) *vs.* wavelength. TDDFT results (blue dots) and fitted $\lambda^{2.85}$ power law (grey line). Except for wavelength, other laser parameters are identical to those in Fig. 1.